\documentclass[aps,twocolumn,nofootinbib,showpacs, prd]{revtex4}
\usepackage{graphicx}
\usepackage{amsfonts}
\usepackage{amssymb}
\usepackage{amsbsy}
\usepackage{amsmath}
\usepackage{latexsym}
\usepackage{natbib}
\usepackage{bm}
\usepackage{subfigure}
\usepackage{color}

\def\ce{\mathrm{ce}}
\def\se{\mathrm{se}}

\def\k{\mathbf{k}}

\def\x{\mathbf{x}}
\def\y{\mathbf{y}}
\def\xiv{\vec{\xi}}

\def\k{\mathrm{k}}

\def\H{\mathcal{H}}

\def\lstar{l_{\star}}

\begin{document}

\title{Transient response from Unruh-DeWitt detector along Rindler trajectory 
in polymer quantization}

\author{Golam Mortuza Hossain}
\email{ghossain@iiserkol.ac.in}

\author{Gopal Sardar}
\email{gopal1109@iiserkol.ac.in}

\affiliation{ Department of Physical Sciences, 
Indian Institute of Science Education and Research Kolkata,
Mohanpur - 741 246, WB, India }
 

\date{\today}

\begin{abstract}

If an Unruh-DeWitt detector operates for an infinite proper time along 
the trajectory of a uniformly accelerated observer in Fock space then induced 
transition rate of the detector is proportional to Planck distribution. For a 
realistic detector which operates only for a finite period, the instantaneous 
transition rate contains both transient and non-transient terms. In particular, 
the non-transient term contains a \emph{residue} evaluated at the pole of the 
two-point function. We show here by considering a massless scalar field that 
unlike in Fock quantization, the short-distance two-point function contains no 
pole in polymer quantization, the quantization techniques used in loop quantum 
gravity. Consequently, corresponding transition rate of the Unruh-DeWitt 
detector contains only transient terms. Thus, the result presented here provides 
an alternative evidence for absence of Unruh effect in polymer quantization 
which was shown earlier by the authors using methods of Bogoliubov 
transformation and Kubo-Martin-Schwinger condition.

\end{abstract}

\maketitle

\section{Introduction}

The Fock vacuum state with respect to a uniformly accelerating observer, 
behaves as a \emph{thermal} state of a given temperature which is proportional 
to the magnitude, say $a$, of the acceleration 4-vector. This phenomena is 
referred as Unruh effect 
\cite{Fulling:1972md,Unruh:1976db,Crispino:2007eb,de2006unruh,Takagi01031986, 
Longhi:2011zj,Davies:1974th}. The corresponding temperature is called 
\emph{Unruh temperature} given by $T = a/2\pi k_B$  where $k_B$ is the 
\emph{Boltzmann constant}. This result follows from the application of standard 
quantum field theory techniques in a curved background 
\cite{Birrell1984quantum}.

The existence of Unruh effect can be seen using many different methods. Firstly, 
one can use the method of \emph{Bogoliubov transformations}. There one computes 
expectation value of number density operator for the accelerated observer in the 
Fock vacuum state. The corresponding expression turns out to be a blackbody 
distribution at Unruh temperature. In this method it appears explicitly that 
Unruh effect depends on the contributions from trans-Planckian modes as observed 
by an \emph{inertial} observer. This aspect makes the Unruh effect to be a 
potentially interesting phenomena to explore and understand the implications of 
possible Planck-scale physics 
\cite{Nicolini:2009dr,Padmanabhan:2009vy,Agullo:2008qb,Chiou:2016exd,Alkofer:2016utc}. In other 
words, the Unruh effect provides a probing window for many candidate theories 
of quantum gravity. Usually these theories predict large modifications for the 
trans-Planckian modes. While this approach is conceptually simpler but in order 
to arrive at the result one is required to use rather sophisticated 
regularization techniques to handle field theoretical divergences. Therefore, 
one is led to pursue other methods of derivation to ensure that the result is 
not an artifact of the employed regularization tools.

An alternative method which is often employed to verify the existence of Unruh 
effect, is to compute the response function of the so-called Unruh-DeWitt 
detector that moves along the trajectory of the accelerated observer 
\cite{DeWitt:1980hx,Hinton:1983dq,Schlicht:2003iy,Louko:2006zv, 
Unruh:1983ms,Louko:2014aba,Sriramkumar:1994pb,Agullo:2010iq,Fewster:2015dqb}. 
This method essentially relies on the concepts of Einstein $A$ and $B$ 
coefficients that are used to describe the \emph{spontaneous} and \emph{induced} 
emission or absorption in statistical physics. In particular, one considers a 
two-level quantum mechanical system as a detector which weakly couples to the 
surrounding matter field. The transition probability between the energy levels 
which is referred as the \emph{response function} of the detector, is then used 
to conclude about the nature of the surrounding environment. It turns out that 
the response function depends on the properties of two-point function of the 
matter field. Given that the Unruh effect receives explicit contributions from 
trans-Planckian modes in the method of Bogoliubov transformations, it is 
compelling to check whether these contributions can also affect the 
corresponding response function. In this article, we compute the response 
function of a Unruh-DeWitt detector in the context of the so called polymer 
quantization of scalar matter field which couples weakly to the detector.

Polymer quantization or loop quantization 
\cite{Ashtekar:2002sn,Halvorson-2004-35} is used as a quantization method in 
loop quantum gravity 
\cite{Ashtekar:2004eh,Rovelli2004quantum,Thiemann2007modern}. It is known to 
differ from the Schr\"odinger quantization method in multiple ways when applied 
to a mechanical system. Firstly, it comes with a new dimension-full parameter 
other than \emph{Planck constant} $\hbar$. In the context of full quantum 
gravity, this new scale essentially corresponds to \emph{Planck length} $l_p$. 
Secondly, in the kinematical Hilbert space which is non-separable, one cannot 
define both position and momentum operators simultaneously but only one of them. 
Due to the non-separability of kinematical Hilbert space, the Stone-von Neumann 
uniqueness theorem is also not applicable. These aspects make polymer 
quantization unitarily \emph{inequivalent} to Schr\"odinger quantization 
\cite{Ashtekar:2002sn}, allowing a different set of results, in principle, from 
polymer quantization.

In the section \ref{Sec:RindlerObserver}, we briefly study the trajectory of  a 
uniformly accelerated observer in Minkowski spacetime and the properties of 
Rindler metric as seen by the accelerated observer. In the section 
\ref{Sec:EinsteinABCoefficients}, we review the basic ideas behind Einstein $A$ 
and $B$ coefficients. In the section \ref{Sec:MasslessScalarField}, we consider 
a massless scalar field in the canonical approach which is convenient for 
application of polymer quantization. We study the properties of the Unruh-DeWitt 
detector in the section \ref{Sec:Unruh-DeWitt-Detector}. Subsequently, we study 
the short and  long distance behaviour of the two-point function in Fock and 
polymer quantization. Then we compute the corresponding response functions. In 
Fock quantization, induced transition rate of the detector contains both 
transient and non-transient terms. In particular, non-transient term is 
proportional to Planck distribution and it contains a residue computed at the 
pole of the corresponding two-point function. On the other hand, we show that 
the two-point function has no pole in polymer quantization. Consequently, the 
corresponding residue which is non-zero in Fock quantization vanishes in polymer 
quantization. So the transition rate in polymer quantization contains only 
transient terms. The result as shown here, provides an alternative evidence to 
the earlier reported results  where it is shown using method of Bogoliubov 
transformation \cite{Hossain:2014fma} as well as using Kubo-Martin-Schwinger 
(KMS) condition \cite{Hossain:2015xqa} that Unruh effect is absent in polymer 
quantization.

%%%%%%%%%%%%%%%%%%%%%

\section{Rindler Observer}\label{Sec:RindlerObserver}

With respect to a uniformly accelerated observer, a section of 
Minkowski spacetime can be described by the so-called Rindler metric. In 
\emph{conformal} Rindler coordinates $\bar{x}^{\alpha} = (\tau,\xi,y,z) \equiv 
(\tau,\xiv)$, the Rindler metric can be expressed as \cite{Rindler:1966zz}
\begin{equation}
 \label{RindlerMetric}
  ds^2 = e^{2a\xi} \left( -d\tau^2 + d\xi^2 \right) + dy^2 + dz^2
  \equiv g_{\alpha\beta}d\bar{x}^{\alpha} d\bar{x}^{\beta}  ~,
\end{equation}
where we have used the \emph{natural} units ($c=\hbar=1$). The parameter $a$ is 
the magnitude of \emph{acceleration} 4-vector. Minkowski metric with 
respect to an \emph{inertial} observer with Cartesian coordinates $x^{\mu} = 
(t,x,y,z) \equiv (t,\x)$ is $ds^2 = \eta_{\mu\nu}dx^{\mu}dx^{\nu} 
= - dt^2 + dx^2 + dy^2 + dz^2$.
The coordinates of Rindler observer \emph{i.e.} the uniformly accelerated 
observer, is related with the coordinates of the inertial observer as 
\begin{equation}
 \label{RindlerMinkowskiRelation}
 t =  \frac{1}{a} e^{a\xi} \sinh a\tau ~,~~~
 x =  \frac{1}{a} e^{a\xi} \cosh a\tau ~,
\end{equation}
where we have assumed that Rindler observer moves along $+ve$ x-axis with 
respect to the inertial observer. Therefore, $y$ and $z$ coordinates are the 
same for both observers. It follows from the equation 
(\ref{RindlerMinkowskiRelation}) that only a wedge-shaped section of Minkowski 
spacetime is covered by Rindler spacetime. The wedge-shaped section is often 
referred as \emph{Rindler wedge}.

\section{Einstein A and B Coefficients}
\label{Sec:EinsteinABCoefficients}

The working principle behind Unruh-DeWitt detector can be understood in 
terms of the so-called \emph{Einstein A and B coefficients} of statistical 
physics. Let us consider a two-level atomic system of energy $E_1$ and $E_2$ 
which is in contact with a \emph{heat bath} of spectral energy density 
$u({\omega})$ such that $\omega \equiv E_2 - E_1 > 0$. The probability of 
transition from the \emph{level $2$} to the \emph{level $1$} \emph{per unit 
time}, \emph{per unit atom} is postulated to be
\begin{equation}
 \label{TwoLevelTransition21}
R_{2\to1} = A + B~ u({\omega}) ~,
\end{equation}
where $A$ and $B$ are the Einstein coefficients. The coefficient $A$ represents 
the transition rate for the \emph{spontaneous} emission whereas the coefficient 
$B$ represents the same for the \emph{induced} emission. On the other hand, 
transition probability \emph{per unit time}, \emph{per unit atom} from the 
\emph{level $1$} to the \emph{level $2$} is postulated to contain only the 
\emph{induced} term as
\begin{equation}
 \label{TwoLevelTransition12}
R_{1\to2} = B~ u({\omega}) ~.
\end{equation}
If the system is in equilibrium with the heat bath  and number of particles in 
the energy \emph{levels $1$, $2$} are $N_1$, $N_2$  respectively, then the 
detailed balance relation implies
\begin{equation}
\label{DetailedBalanceRelation}
N_1 R_{1\to2} = N_2 R_{2\to1} ~.
\end{equation}
On the other hand, particles numbers would satisfy Boltzmann distribution law 
\emph{i.e.} $N_2/N_1 = \exp(-\omega/k_B T)$. So if the ratio of the Einstein 
coefficients is taken to be $A/B = \omega^3/\pi^2$, then it precisely leads to 
Planck distribution formula for spectral energy density of blackbody radiation
\begin{equation}
\label{PlanckDistributionDensity}
u({\omega}) = \frac{\omega^3}{\pi^2} \frac{1}{e^{\omega/k_B T} - 1} 
~.
\end{equation}
Clearly, the induced transition rate of the two-level system in equilibrium can 
be used to conclude about the thermal nature of the surrounding environment. In 
particular, the equation (\ref{TwoLevelTransition12}) implies that if the 
induced transition rate from the \emph{level $1$} to the \emph{level $2$} is 
proportional to Planck formula for  \emph{mean energy} $\epsilon_{\omega}$ 
of a given mode with angular frequency $\omega$ \emph{i.e.}
\begin{equation}
\label{TwoLevelTransition12NumberDensity}
R_{1\to2} ~\propto~ \epsilon_{\omega} =  \frac{\omega}{e^{\omega/k_B T} - 1} ~.
\end{equation}
then surrounding environment must be a thermal radiation at a temperature $T$. 
We note that only the ratio of the Einstein coefficients is relevant here but 
not their absolute values.

\section{Massless scalar field}
\label{Sec:MasslessScalarField}

In order to study the response function of the Unruh-DeWitt detector, we 
consider here a massless scalar field $\Phi(x)$ in Minkowski spacetime
that weakly couples to the detector. The corresponding scalar field
dynamics is described by the action
\begin{equation}
 \label{ScalarActionMinkowski}
 S_{\Phi} = \int d^4x \left[ - \frac{1}{2} \sqrt{-\eta} \eta^{\mu\nu}  
 \partial_{\mu} \Phi(x)  \partial_{\nu} \Phi(x) \right] ~.
\end{equation}
In the canonical formulation, the dynamics of the scalar field can be 
equivalently described by the field Hamiltonian 
\begin{equation}\label{SFHamGen}
H_{\Phi}  =  \int d^3\x \left[ \frac{\Pi^2}{2\sqrt{q}} +
\frac{\sqrt{q}}{2} q^{ab} \partial_a\Phi \partial_b\Phi
\right] ~,
\end{equation}
where $q_{ab}$ is the metric on \emph{spatial} hyper-surfaces labeled by $t$. 
Poisson bracket between the field $\Phi = \Phi(t,\x)$ and its conjugate 
field momentum $\Pi = \Pi(t,\x)$ is 
\begin{equation}\label{PositionSpacePB}
\{\Phi(t,\x), \Pi(t,\y)\} = \delta^{3}(\x-\y) ~,
\end{equation}
where $\delta^{3}(\x-\y)$ is the Dirac delta.

\subsection{Fourier modes}

It is convenient to use Fourier space for studying dynamics of a massless free 
scalar field. Here we define Fourier modes for the scalar field and its 
conjugate field momentum as
\begin{equation}\label{FourierModesDef}
\Phi = \frac{1}{\sqrt{V}} \sum_{\k} \tilde{\phi}_{\k}(t) e^{i \k\cdot\x} ,~
\Pi  = \frac{1}{\sqrt{V}} \sum_{\k} \sqrt{q} ~\tilde{\pi}_{\k}(t) 
e^{i \k\cdot\x},
\end{equation}
where $V=\int d^3\x \sqrt{q}$ is the spatial volume. The space being 
non-compact, the spatial volume would normally diverge for Minkowski spacetime. 
Therefore, it is often convenient to use a fiducial box of finite volume so that 
one can avoid dealing with such divergent quantity. Kronecker delta and Dirac 
delta then can be expressed as $\int d^3\x \sqrt{q} ~e^{i (\k-\k')\cdot \x} = V 
\delta_{\k,\k'}$ and $\sum_{\k} e^{i \k\cdot (\x-\y)} = V \delta^3 
(\x-\y)/\sqrt{q}$. 
The field Hamiltonian (\ref{SFHamGen}) can be expressed as $H_{\Phi} = 
\sum_{\k} \H_{\k}$, where Hamiltonian density for the $\k-$th Fourier mode is
\begin{equation}\label{SFHamFourierMinkowski}
\H_{\k} = \frac{1}{2} \tilde{\pi}_{-\k} \tilde{\pi}_{\k} +
\frac{1}{2} |\k|^2 \tilde{\phi}_{-\k}\tilde{\phi}_{\k}  ~.
\end{equation}
Poisson brackets between these Fourier modes and their conjugate momenta 
are given by
\begin{equation}\label{Minkowski:MomentumSpacePB}
\{\tilde{\phi}_{\k}, \tilde{\pi}_{-\k'}\} =  \delta_{\k,\k'}
~.
\end{equation}
In order to satisfy the \emph{reality condition} of the scalar field $\Phi$, 
one usually redefines the \emph{complex-valued} modes $\tilde{\phi}_{\k}$ 
and momenta $\tilde{\pi}_{\k}$ in terms of the real-valued functions 
$\phi_{\k}$ and $\pi_{\k}$ such that corresponding Hamiltonian density and 
Poisson brackets become
\begin{equation}\label{SFHamFourierMinkowskiReal}
\H_{\k} = \frac{1}{2} \pi_{\k}^2 + \frac{1}{2} |\k|^2 \phi_{\k}^2
~~~;~~~ \{\phi_{\k},\pi_{\k'}\} =  \delta_{\k,\k'} ~.
\end{equation}
This is the standard Hamiltonian for a system of decoupled harmonic 
oscillators. We may express the energy spectrum of these Fourier modes as 
$\hat{\H}_{\k}|n_{\k}\rangle = E_n^{(\k)}|n_{\k}\rangle$. Using the vacuum state 
of each mode $|0_\k\rangle$, we can express the vacuum state of the scalar field 
as $|0\rangle=\Pi_{\k}\otimes |0_\k\rangle$.

\section{Unruh-DeWitt Detector}
\label{Sec:Unruh-DeWitt-Detector}

Unruh-Dewitt detector is considered to be a point-like quantum mechanical 
system having two internal energy levels. These system of detectors interact 
\emph{weakly} with the scalar field through a \emph{linear} coupling which is 
treated as perturbative interaction to the detectors. If we denote the 
Hamiltonian operator of an \emph{unperturbed} detector by $\hat{H}_0$ then we 
may express the energy eigenvalue equation of the detector as
\begin{equation}\label{DetectorEigenvalueEquation}
\hat{H}_0 |g\rangle = \omega_g|g\rangle ~~;~~
\hat{H}_0 |e\rangle = \omega_e|e\rangle ~,
\end{equation}
where $|g\rangle$ and $|e\rangle$ represent the \emph{ground state} and the
\emph{excited state} respectively. We may denote the energy gap between the 
levels as
\begin{equation}\label{DetectorEnergyGap}
\omega \equiv \left(\omega_e - \omega_g \right) > 0 ~.
\end{equation}
The interaction term in the Hamiltonian of Unruh-DeWitt detector is taken to be 
of the form
\begin{equation}\label{DetectorFieldInteractionHamiltonian}
\hat{H}_{int} (\tau) = \lambda ~\hat{m}(\tau) \hat{\Phi}(x(\tau)) ~,
\end{equation}
where $\lambda$ denotes the \emph{coupling constant} and $\hat{m}(\tau)$ is 
the monopole moment operator of the detector. The world line  $x^{\mu}(\tau)$ 
of the detector is parametrized using the proper time $\tau$. Therefore, the 
total Hamiltonian of the detector becomes 
\begin{equation}\label{TotalSystemHamiltonian}
\hat{H} = \hat{H}_0 + \hat{H}_{int}  ~.
\end{equation}
Given the nature of the Hamiltonian, it is convenient to work in the so-called 
`interaction picture' of quantum mechanics. In particular, by defining 
interaction state $|\psi_{\tau}\rangle_I$ in terms of Schr\"odinger state 
$|\psi_{\tau}\rangle_S$ as
\begin{equation}\label{SchrodingerToInteraction}
|\psi_{\tau}\rangle_I \equiv e^{i \hat{H}_0 \tau } |\psi_{\tau}\rangle_S 
~~;~~
\hat{H}_I \equiv e^{i \hat{H}_0 \tau} \hat{H}_{int} e^{-iH_0 \tau} ~,
\end{equation}
one can express the time evolution of the detector state as 
$|\psi_{\tau_f}\rangle_I = U(\tau_f,\tau_i) |\psi_{\tau_i}\rangle_I$
where the evolution operator is given by
\begin{equation}\label{EvolutionOperator}
U(\tau_f,\tau_i) =  1 - i \int_{\tau_i}^{\tau_f} d\tau' U(\tau',\tau_i)
\hat{H}_I (\tau') ~.
\end{equation}
We may denote the combined state of the detector and the scalar 
field at a given proper time $\tau$ as
\begin{equation}\label{GeneralCombinedState}
|\psi, \Theta; \tau\rangle \equiv |\psi_{\tau}\rangle_I \otimes
|\Theta_{\tau}\rangle  ~,
\end{equation}
where $|\Theta_{\tau}\rangle$ denotes the state for the scalar field. 
Therefore, the transition \emph{amplitude} for the combined system 
going from the state $|g, \Theta_i; 0\rangle$ to the 
$|e, \Theta_f; \tau\rangle$ can be written as
\begin{equation}\label{TransitionAmplitude}
\mathcal{A} = - i \lambda \int_{0}^{\tau} d\tau' 
\langle e, \Theta_f; \tau| \hat{m}_I(\tau') \hat{\Phi}(x(\tau')) 
|g, \Theta_i;0\rangle ~,
\end{equation}
where $\hat{m}_I(\tau)$ is the monopole moment operator in the interaction 
picture. The corresponding \emph{probability} of transition then becomes
\begin{equation}\label{TransitionProbability}
P_{|g, \Theta_i; 0\rangle \to |e, \Theta_f; \tau\rangle} = |\mathcal{A} |^2 ~.
\end{equation}
We are interested here in considering the induced transition of the detector 
which is initially at the ground state $|g\rangle$ and the scalar field is in 
its vacuum state \emph{i.e.} $|\Theta_i\rangle = |0\rangle$. Then transition 
probability of the detector being at the excited state $|e\rangle$ at a later 
time $\tau$ for all possible field state is
\begin{equation}\label{TransitionProbabilityDetector}
P_{\omega}(\tau,0) \equiv P_{|g; 0\rangle \to |e; \tau\rangle} = 
\sum_{\{|\Theta_f\rangle\}} P_{|g, \Theta_i; 0\rangle \to |e, \Theta_f; 
\tau\rangle} ~.
\end{equation}
It is now straightforward to express the transition probability 
(\ref{TransitionProbabilityDetector}) in the form of
\begin{equation}\label{TransitionProbabilityDetector2}
P_{\omega}(\tau,0) = A_0 F_{\omega}(\tau,0) ~,
\end{equation}
where $A_0 = \lambda^2 |\langle e| \hat{m}(0) |g\rangle|^2$. $A_0$ depends on 
the internal structure of the detector system. The function $F_{\omega}(\tau,0)$ 
is known as the \emph{response function} of the detector and is defined as
\begin{equation}\label{DetectorResponseFunction}
F_{\omega}(\tau,0) = \int_{0}^{\tau} \int_{0}^{\tau} d\tau' d\tau''
e^{-i\omega(\tau''-\tau')} ~G(\tau'',\tau') ~,
\end{equation}
where $G(\tau'',\tau')$ is the \emph{two-point function} of the scalar field
and is given by
\begin{equation}\label{TwoPointFunction}
G(\tau'',\tau') = G(\tau''-\tau')  =
\langle 0| \hat{\Phi}(x(\tau'')) \hat{\Phi}(x(\tau')) |0\rangle ~.
\end{equation}

We may recall that only the ratio of the Einstein coefficients can be 
determined but not their absolute values. We shall use this freedom to scale 
the response function of the detector such that proportionality constant 
between the induced transition rate and the mean energy per mode,
as in equation (\ref{TwoLevelTransition12NumberDensity}), becomes identity.
Using equation (\ref{TransitionProbabilityDetector2}) we define the
\emph{instantaneous} transition rate, after being scaled by the chosen factor, 
as
\begin{equation}\label{DetectorTransitionRate}
R_{\omega}(\tau,0) \equiv \left(\frac{2\pi}{A_0} \right)
\frac{d P_{\omega}}{d\tau}
= 2\pi \int_{-\tau}^{\tau} d\tau' e^{-i\omega \tau'} G(\tau') ~,
\end{equation}
which can be further expressed as
\begin{equation}\label{DetectorTransitionRate2  }
R_{\omega}(\tau,0) = R_{\omega}^0 + \Delta R_{\omega}(\tau) ~.
\end{equation}
Here $R_{\omega}^0$ represents time independent \emph{i.e.} 
\emph{non-transient} 
part of the induced transition rate and its expression is given by
\begin{equation} \label{IntegralR0}
R_{\omega}^0 =  2\pi \int_{-\infty}^{\infty} d\tau' 
e^{-i\omega \tau'} G(\tau') ~.
\end{equation}
On the other hand $\Delta R_{\omega}(\tau)$ represent time-dependent \emph{i.e.} 
\emph{transient} part of the induced transition rate and its expression is
\begin{equation} \label{IntegralDeltaR}
\Delta R_{\omega}(\tau) = - 2\pi \int_{\tau}^{\infty} d\tau' 
\left[ e^{-i\omega \tau'} G(\tau') +  e^{i\omega \tau'} G(-\tau') \right] ~.
\end{equation}
Clearly, $\Delta R_{\omega}(\tau) \to 0$  as the time of observation  
increases \emph{i.e.} $\tau \to \infty$. 

We should emphasize here that the definition of instantaneous transition rate 
(\ref{DetectorTransitionRate}) presumes the so-called \emph{sharp switching} 
of the Unruh-DeWitt detector. While sharp switching of the detector is used 
quite widely in the literature, it has certain key weaknesses. This issue has 
been studied in \cite{Satz:2006kb,Langlois:2005nf,Louko:2006zv} and it is shown 
that one could overcome these weaknesses by considering either a spatially 
extended detector or a smooth switching function without affecting the main 
result regarding Unruh effect.

\section{Two-point function}\label{sec:two-point-function}

It is evident from the equation (\ref{DetectorTransitionRate}) that 
induced transition rate of the Unruh-DeWitt detector is fully determined once 
the properties of the two-point function is known. In terms of the Fourier 
modes (\ref{FourierModesDef}), the general form of such a two-point function can 
be written as
\begin{equation}
\label{MinkowskiTwoPointDef2}
G(x,x') = \langle 0|\hat{\Phi}(x) \hat{\Phi}(x')|0\rangle =
\frac{1}{V} \sum_{\k} D_{\k}(t,t') e^{i {\k}\cdot(\x-\x')} ,
\end{equation}
where the matrix element $D_{\k}(t,t')$ is given by
\begin{equation}\label{DkDefinition}
D_{\k}(t,t') = 
\langle 0_{\k}| e^{i\hat{\H}_{\k}t} \hat{\phi}_{\k} e^{-i\hat{\H}_{\k}t}
e^{i\hat{\H}_{\k}t'} \hat{\phi}_{\k} e^{-i\hat{\H}_{\k}t'}
|0_{\k}\rangle.
\end{equation}
We should note that due to the chosen definition of Fourier modes 
(\ref{FourierModesDef}), the Hamiltonians and the corresponding Poisson 
brackets (\ref{SFHamFourierMinkowskiReal}) are independent of the fiducial 
volume. Therefore, we can easily remove the fiducial box by taking the limit $V 
\to \infty$. This essentially replaces the sum $\frac{1}{V} \sum_{\k}$ by an 
integration $\int \frac{d^3\k}{(2\pi)^3}$
in the expression of the two-point function (\ref{MinkowskiTwoPointDef2}) as
\begin{equation}
\label{MinkowskiTwoPointIntegralDef}
G(x,x') = \int \frac{d^3\k}{(2\pi)^3} ~  D_{\k}(t,t') 
~e^{i {\k}\cdot(\x-\x')} ~.
\end{equation}
Using energy spectrum of the Fourier modes and by expanding the state 
$\hat{\phi}_{\k} |0_{\k}\rangle$ in the basis of energy eigenstates as 
$\hat{\phi}_{\k}|0_{\k}\rangle = \sum_{n} c_n |n_{\k}\rangle$, the matrix 
element $D_{\k}(t,t')$ formally becomes
\begin{equation}
\label{DkFunctionGeneral}
D_{\k}(t-t') \equiv D_{\k}(t,t') = \sum_{n} |c_n|^2 e^{-i\Delta E_n (t-t')},
\end{equation}
where $\Delta E_n \equiv E_n^{(\k)} - E_0^{(\k)}$ and
$c_n = \langle n_{\k}| \hat{\phi}_{\k} |0_{\k}\rangle$. Given the matrix 
element $D_{\k}(t-t')$ depends only on magnitude $|\k|$, one can carry
out the angular integration using \emph{polar coordinates}, to reduce the 
two-point function (\ref{MinkowskiTwoPointIntegralDef}) as
\begin{equation}
\label{KGPropagatorDiffPM}
G(x,x') = G_{+} - G_{-} ~,
\end{equation}
where
\begin{equation}
\label{GPMDefinition}
G_{\pm} =  \frac{i}{4\pi^2|\Delta\x|} \int dk ~k D_{k}(\Delta t) 
~e^{\mp i k |\Delta \x|} ~,
\end{equation}
with $k = |\k|$, $\Delta \x = \x-\x'$ and $\Delta t = t-t'$ 
\cite{Hossain:2015xqa}.

\section{Fock quantization}

In Fock quantization of the scalar field, Fourier modes which are represented by 
the equation (\ref{SFHamFourierMinkowskiReal}), are essentially quantized using 
Schr\"odinger quantization method. The corresponding energy spectrum and the 
coefficients $c_n$ (\ref{DkFunctionGeneral}) are given by
\begin{equation}\label{FockSpectrum}
E^n_{\k} = \left(n+\frac{1}{2}\right)|\k| ~~;~~ 
\Delta E_n = n |\k|   ~~;~~
 c_n = \frac{\delta_{1,n}}{\sqrt{2|\k|}} ~.
\end{equation}
The two-point function (\ref{KGPropagatorDiffPM}) then reduces to its standard 
form for the Fock space 
\begin{equation}
\label{FockTwoPointFunction}
G(x,x') = \frac{1}{4\pi^2 \left[-(\Delta t - i\delta)^2 
+ |\Delta \x|^2\right]}~,
\end{equation}
where $\Delta x^2 = - \Delta t^2 + |\Delta \x|^2$ is Lorentz invariant 
spacetime interval and $\delta$ is a small, positive parameter that is 
introduced as the standard integral regulator.

\subsection{Detector response along inertial trajectory}

In order to illustrate the nature of the \emph{transient} response of the 
Unruh-DeWitt detector, for simplicity, we consider a detector which is 
located at a fixed position in an inertial frame \emph{i.e.} its world line is 
given by $x_d^{\mu}(\tau) = (\tau,x_0,y_0,z_0)$ where proper time $\tau$ is 
defined as $\Delta \tau^2 = - \Delta x^2$. Along the world line 
of the detector the two-point function becomes 
\begin{equation}\label{FockTwoPointFunctionIntertial}
G(\tau) \equiv G(x_d(\tau),x_d(0)) = - \frac{1}{4\pi^2 (\tau - 
i\delta)^2} ~.
\end{equation}
%
%where the redefined small parameter $\delta = \epsilon \tau/2 > 0$ also acts 
%as an integral regulator. 
One would get the same two-point function even for a 
detector which moves with a uniform velocity.

We can compute the non-transient part of the transition rate $R_{\omega}^0$ 
(\ref{IntegralR0}) by closing the contour in the lower half of 
the complex plane. We note that the two-point function 
(\ref{FockTwoPointFunctionIntertial}) has a \emph{pole} at $\tau = i \delta$
and the contour does \emph{not} enclose the pole. Therefore, the non-transient 
part of the transition rate vanishes \emph{i.e.}
\begin{equation}\label{IntertialFrameR0}
R_{\omega}^0 = 0 ~. 
\end{equation}
The \emph{transient} part $\Delta R_{\omega}(\tau)$ can also be computed in a 
straightforward manner as
\begin{equation}\label{DetectorResponseFunctionFockInertial}
\Delta R_{\omega}(\tau) = \frac{\omega}{\pi} 
\left[\frac{\cos(\omega\tau)}{\omega\tau} + 
\left\{ \mathrm{Si}(\omega\tau) - \frac{\pi}{2} \right\} \right]  ~,
\end{equation}
where $\mathrm{Si}(\omega\tau)$ is the \emph{sine integral} function which goes 
to $\pi/2$ as $\tau \to \infty$. 
Clearly, the induced transition rate of an Unruh-DeWitt detector is 
\emph{non-vanishing} even along an inertial trajectory if the observation is 
made for a finite period. However, as expected, such a response of the detector 
decays out as the time of observation increases. Therefore, in order to conclude 
about \emph{thermal} nature of the surrounding environment, one must look at 
the \emph{non-transient} part of induced transition rate of the detector.

\subsection{Detector response along Rindler trajectory}

We now consider an Unruh-DeWitt detector which moves along a Rindler 
trajectory given by $x_d^\mu (\tau) = (\sinh(a\tau)/a,\cosh(a\tau)/a,0,0)$. For 
convenience, we define a new variable $\eta = e^{a \tau}$. It leads 
the time interval $\Delta t$ and spatial separation $|\Delta \x|$ to become
\begin{equation}\label{RindlerDeltaXandT}
\Delta t = \frac{(\eta^2-1)}{2 a\eta} ~~,~~
|\Delta \x| = \frac{(\eta-1)^2}{2 a\eta} ~,
\end{equation}
and the corresponding two-point function to become
\begin{equation}\label{FockTwoPointFunctionRindlerEta}
G(\eta) =- \frac{a^2~ \eta}
{4 \pi^2 \left(\eta - 1 -i\delta \right)^2}  ~.
\end{equation}
The non-transient part of the transition rate (\ref{IntegralR0}) then becomes
\begin{equation}\label{IntegralR0FockRindler}
R_{\omega}^0 = \frac{2\pi}{a} \int_{0}^{\infty} d\eta
~\eta^{-1 - i\omega/a} G(\eta) 
\equiv \int_{0}^{\infty} d\eta f(\eta) ~,
\end{equation}
which may also be expressed as
\begin{equation}\label{IntegralR0Fock3}
R_{\omega}^0 = -\frac{a}{2\pi} \int_{0}^{\infty} d\eta 
\frac{\eta^{-i\omega/a}} {\left(\eta - 1 - i\delta\right)^2} ~.
\end{equation}
We note that integrand $f(\eta)$ has a \emph{pole of second order} at 
$\eta=1+i\delta$ and it is a \emph{multi-valued} function of complex variable 
$\eta$. 
\begin{figure}
 \includegraphics[scale=.5]{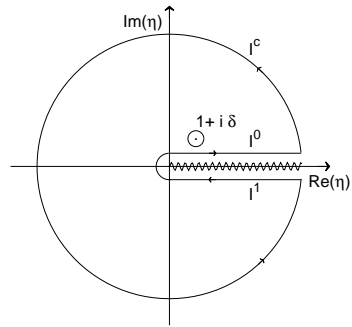}
 \caption{\label{fig:contour}Contour used to evaluate $R_{\omega}^0$}
\end{figure}
Following the contour in the complex plane as shown in FIG.~\ref{fig:contour}, 
we express the contour integral as
\begin{equation}\label{IntegralI0FockRindler}
\oint d\eta f(\eta) = 
I^0 + I^1 + I^c = (2\pi i)~
\mathrm{Res}[f(\eta)]_{|\eta=1+i\delta} ~,
\end{equation}
where $\mathrm{Res}[f(\eta)]_{|\eta=1+i\delta}$ denotes the \emph{residue} of 
the function $f(\eta)$ evaluated at the pole $\eta=(1+i\delta)$. Using the 
properties that $I^c =0$ and $I^1 = - e^{2\pi\omega/a}~ I^0$, non-transient part 
of the transition rate can be expressed as
\begin{equation}\label{RindlerGeneralR0}
R_{\omega}^0  = I^0 =  
\frac{-(2\pi i)\mathrm{Res}[f(\eta)]_{|\eta=1+i\delta}}{e^{2\pi\omega/a} - 1} ~.
\end{equation}
The evaluated residue at the pole of the two-point function in Fock space
\begin{equation}\label{FockRindlerResidue}
\mathrm{Res}[f(\eta)]_{|\eta=1+i\delta} = \frac{i \omega}{2\pi}  ~,
\end{equation}
leads the induced transition rate to become
\begin{equation}\label{RindlerFockR0Final}
R_{\omega}^0  = \frac{\omega}{e^{2\pi\omega/a} - 1} ~.
\end{equation}
In other words, the induced transition rate $R_{\omega}^0$ is precisely equal 
to the standard expression for mean energy per mode of a system in thermal 
equilibrium at the Unruh temperature $T = a/2\pi k_B$.

The transient part $\Delta R_{\omega}$ can be computed easily
by considering reasonably large but finite time of observation as
\begin{equation}\label{IntegralDeltaIFockRindler}
\Delta R_{\omega}({\tau}) \approx \frac{e^{-a\tau}}
{\pi \left( 1+ \omega^2/a^2 \right)} 
\left[a \cos(\omega \tau) -\omega \sin(\omega\tau)\right].
\end{equation}
The transient terms decays out exponentially as the time of observation 
increases.

We summarize two key inputs that are required to arrive at the equation 
(\ref{RindlerFockR0Final}). Firstly, to ensure $I^c = 0$ the function $f(\eta)$ 
should have sufficient fall-off as $\eta\to\infty$ so that \emph{Jordan's lemma} 
is applicable \emph{i.e.} $\lim_{\eta\to\infty} |f(\eta)| = 
\lim_{\eta\to\infty} |G(\eta)/\eta| = 0$. 
Secondly, in order to have a \emph{non-zero} residue, the function 
$f(\eta)$ should have at least one \emph{pole} within the region enclosed by the 
contour. It is evident from the equation (\ref{IntegralR0Fock3}), the 
\emph{short-distance} singular nature of the Fock space two-point function 
is a key requirement to ensure non-vanishing residue. In other words, the pole 
of two-point function $G(\eta)$ at ($\eta - 1) = i\delta$ plays the crucial 
role for the existence of the Unruh effect.

\section{Polymer quantization}

In order to perform the so-called polymer quantization of scalar field, we 
follow the approach as suggested here \cite{Hossain:2010eb} and was followed up 
later to study Unruh effect \cite{Hossain:2014fma,Hossain:2015xqa}. In 
this approach, one quantizes the system of Fourier modes using polymer 
quantization instead of Schr\"odinger quantization as used in Fock space. 
As mentioned earlier, Polymer quantization or \emph{loop quantization} is a 
quantization technique that is used in \emph{loop quantum gravity} and it comes 
with a new dimension-full parameter say $\lstar$ along with Planck constant 
$\hbar$. In full quantum gravity, the parameter $\lstar$ would be analogous to 
Planck length. 

The energy spectrum of the $\k-$th oscillator mode in polymer 
quantization are given by \cite{Hossain:2010eb}  
\begin{equation}
 \label{EigenValueMCFRelation}
 \frac{E_{\k}^{2n}}{|\k|} = \frac{1}{4g} + \frac{g}{2} ~A_n(g)  ~,~
 \frac{E_{\k}^{2n+1}}{|\k|} = \frac{1}{4g} + \frac{g}{2} B_{n+1}(g) ~,
\end{equation}
where $n\ge0$, $A_n$, $B_n$ are \emph{Mathieu characteristic value
functions} and $g = |\k|~\lstar$ is a \emph{dimension-less} 
parameter. The energy eigenstates are 
$\psi_{2n}(v) = \ce_n(1/4g^2,v)/\sqrt{\pi}$ and 
$\psi_{2n+1}(v) = \se_{n+1}(1/4g^2,v)/\sqrt{\pi}$ where 
$v = \pi_{\k} \sqrt{\lstar} + \pi/2 $.
The functions $\ce_n$ and $\se_n$ are solutions to \emph{Mathieu
equation}. They are referred as elliptic cosine and sine functions respectively 
\cite{Abramowitz1964handbook}. In order to arrive at these $\pi$-periodic and 
$\pi$-antiperiodic states in $v$, superselection rules are invoked. This 
superselection leads to exact energy spectrum which in turns allows to study the 
system analytically. Besides, without imposition of superselection rules, 
certain key statistical notions for the system are known to become ill-defined 
\cite{Barbero:2013lia}.

For low-energy modes \emph{i.e.} for small $g$, the energy spectrum
(\ref{EigenValueMCFRelation}) reduces to regular harmonic oscillator energy 
spectrum along with perturbative corrections 
\begin{equation}
 \label{EEvalueSmallg}
  \frac{E_{\k}^{2n}}{|\k|} \approx \frac{E_{\k}^{2n+1}}{|\k|} \approx
   \left(n+\frac{1}{2}\right) + \mathcal{O}(g)~.
\end{equation}
Therefore, polymer quantization leads to expected results for low-energy modes. 
However, we note that polymer energy spectrum has two-fold degeneracy as $g\to0$ 
and it is lifted for finite values of $g$. The coefficients  $c_{4n+3} = i 
\sqrt{\lstar} \int_0^{2\pi} \psi_{4n+3} \partial_v\psi_{0} dv$ are non-vanishing 
in polymer quantization for all $n=0, 1, 2, \ldots$, unlike in Fock quantization 
where only \emph{one} $c_n$ is non-vanishing (\ref{FockSpectrum}). 
Using asymptotic properties of Mathieu functions, we can approximate
the energy gaps $\Delta E_n$ between the levels and coefficients $c_{4n+3}$ for 
low-energy modes or \emph{sub-Planckian modes} (\emph{i.e.} $g\ll 1$) as
\begin{equation}
\frac{\Delta E_{4n+3}}{|\k|} = (2n+1) - \frac{(4n+3)^2-1}{16} g  + 
\mathcal{O} \left( g^2 \right) ~,
\end{equation}
for $n \ge 0$, and 
\begin{equation}
\label{SmallgDeltaE4n+3}
  c_3 = \frac{i}{\sqrt{2|\k|}} \left[1 
  +\mathcal{O}\left(g\right) \right] ~,~
  \frac{c_{4n+3}}{c_{3}} = \mathcal{O}\left(g^n\right),
\end{equation}
for $n >0$. On the other hand, for high energy modes or \emph{super-Planckian} 
modes (\emph{i.e.} $g\gg1$), we can approximate the energy gaps and  
coefficients $c_{4n+3}$ as
\begin{equation}
 \label{LargegDeltaE4n+3}
 \frac{\Delta E_{4n+3} }{|\k|}  =  2(n+1)^2 g + 
  \mathcal{O}\left(\frac{1}{g^3}\right) ,
\end{equation}
for $n \ge 0$, and 
\begin{equation}
  \label{LargegC4n+3}
  c_{3} =  i\sqrt{\frac{g}{2|\k|}} \left[\frac{1}{4g^2} +
  \mathcal{O}\left(\frac{1}{g^6}\right) \right],
  \frac{c_{4n+3}}{c_{3}} = \mathcal{O}\left(\frac{1}{g^{2n}}\right),
\end{equation}
for $n > 0$. Therefore, we can approximate matrix element $D_{\k}(\Delta 
t)$ in polymer quantization as
\begin{equation}
\label{DkPolyApprox1}
D_{\k}^{poly}(\Delta t) \simeq  |c_3|^2 e^{ -i \Delta E_3 \Delta t} ~,
\end{equation}
for both the cases.

\subsection{Short-distance two-point function}

As discussed earlier, the singular nature of the short-distance two-point 
function plays the key role in providing non-vanishing residue to the expression 
of induced transition rate (\ref{RindlerFockR0Final}) in Fock space. Therefore, 
in order to understand the effect of polymer quantization on the response 
function of Unruh-DeWitt detector it is crucial to determine the form of 
short-distance two-point function (\emph{i.e.} near $(\eta -1) = 0$) in polymer 
quantization.

Given $\eta \to 1$ implies $|\Delta \x|\to 0$,  we may express the 
two-point function $G(x,x') = G_{+} - G_{-}$, as a series of the form
\begin{equation}\label{TwoPointFunctionSeriesGeneral}
G = \sum_{m=1}^{\infty}\frac{(-2 i |\Delta \x|)^{m-1}}{2\pi^2 m!} 
\int_0^{\infty} dk~k^{m+1} D_{k}(\Delta t)e^{i\k|\Delta \x|} ~,
\end{equation}
where we have used the identity $e^{-x} = e^x \sum_{m=0}^{\infty} 
(-2x)^m/m!$.

In Fock quantization, the matrix element has exact expression $D_{\k}(\Delta t) 
= (1/2|\k|) e^{ -i |\k| \Delta t}$ whereas in polymer quantization we can 
approximate it as $D_{\k}(\Delta t) \simeq  |c_3|^2 e^{ -i \Delta E_3 \Delta 
t}$. In other words, we can represent the matrix element for both the cases in 
a general form $D_{\k}(\Delta t) =  |c_k|^2 e^{ - i\Delta E_k \Delta t}$. 
Further, in the domain where $(\eta - 1)$ is very small, the temporal and the 
spatial separations satisfies $\Delta t \gg |\Delta \x|$ and $\Delta E_3/k$ is 
at least $\mathcal{O}(1)$ or higher for all modes. So by defining a new 
variable $u=\Delta E_k(\eta-1)/a$ and keeping only the leading term, we may 
approximate the two-point  function (\ref{TwoPointFunctionSeriesGeneral}) as
\begin{equation}\label{TwoPointFunctionPolymerFirstOrder4}
G(\eta) = \int_0^{\infty} du ~h(u,\eta) ~ e^{-iu} 
+ \mathcal{O}(\eta-1) ~.
\end{equation}
where
\begin{equation}\label{hGeneral}
h(u,\eta) = \frac{k^2|c_k|^2}{2\pi^2} \frac{dk}{du} ~.
\end{equation}
Unlike in Fock quantization, the expression for $|c_k|^2$ and 
$\Delta E_k$ are different for sub-Planckian and super-Planckian 
modes in polymer quantization due to the presence of the scale $\lstar$.
Therefore, here we consider the sub-Planckian and super-Planckian 
contributions to the two-point function separately as
\begin{equation}
 G(\eta) = G_{sub}(\eta) + G_{super}(\eta).
\end{equation}
The sub-Planckian contributions to the two-point function, is defined as
\begin{equation}\label{GSub}
G_{sub}(\eta) \equiv \int_0^{u_0} du ~h(u,\eta)~e^{-iu} 
+ \mathcal{O}(\eta-1) ~,
\end{equation}
where $u_0 = \Delta E_{k_0}(\eta-1)/a$ and $k_0$ is some pivotal value of 
the wave-vector chosen such that $k_0 \lstar$ is $\mathcal{O}(1)$. The limit
$\eta\to1$ implies $u_0\to0$ which in turns leads $G_{sub}(\eta)\to
\mathcal{O}(\lstar^{-2})$. On the other hand, the super-Planckian contributions 
to the two-point function, defined by
\begin{equation}\label{GSuper}
G_{super}(\eta) = \int_{u_0}^{\infty} du~h(u,\eta)~ e^{-iu} ~,
\end{equation}
are expected to dictate the short-distance ($\eta \to 1$) behaviour of the 
two-point function. Using the asymptotic expressions (\ref{LargegDeltaE4n+3}) 
and (\ref{LargegC4n+3}), one can compute the expression for $h(u,\eta)$ for 
polymer quantization as
\begin{equation}\label{hPolySuper}
h^{poly}(u,\eta) = \frac{u^{-3/2}}{64\pi^2 \lstar^2} 
\sqrt{\frac{\eta-1}{2a\lstar}} 
\left[1 + \mathcal{O}(\eta-1)\right] ~.
\end{equation}
Therefore, the short-distance two-point function with \emph{non-perturbative}
modifications from polymer quantization can be expressed as
\begin{equation}\label{GPolyFinal}
 G^{poly}(\eta) =  -\frac{(1+i)}{64\pi \lstar^2} 
\sqrt{\frac{\eta-1}{\pi a\lstar}} \left[1 + \mathcal{O}(\eta-1)\right]
+\mathcal{O}(\lstar^{-2}) ~,
\end{equation}
where we have used \emph{analytic continuation} to evaluate the integral as
$\int_0^{\infty} du ~u^{-3/2} e^{-iu} = - \sqrt{2\pi} (1+i)$.

We note here that the short-distance two-point function (\ref{GPolyFinal}) 
instead of diverging, it reaches a maximum value of $\mathcal{O}(\lstar^{-2})$. 
This bounded from above behaviour of the two-point function is analogous 
to the behaviour of the spectrum of \emph{inverse scale factor} operator as well 
as the effective Hubble parameter in loop quantum cosmology (LQC) 
\cite{Bojowald:2001vw,Bojowald:2001xe,Ashtekar:2006rx,Ashtekar:2006wn}. Both of 
these can be associated with some inverse powers of the distance similar to the 
two-point function. In LQC, this crucial behaviour plays a key role in 
resolution of Big Bang singularity. However unlike in LQC, here we have applied 
polymer quantization only for scalar matter field rather than for the geometry.

We may verify that the equation (\ref{GSuper}) also reproduces the result of 
Fock quantization up to $\mathcal{O}(\eta-1)$ term as 
\begin{equation}\label{TwoPointFunctionFockFirstOrder2}
G(\eta) = - \frac{a^2}{4\pi^2 (\eta-1 -i\delta)^2} 
\left[1 + \mathcal{O}(\eta-1)\right] ~.
\end{equation}
Here we have used the Fock space expression $h(u,\eta) = 
\left(a/2\pi(\eta-1)\right)^2 u$ that follows from the equation 
(\ref{FockSpectrum}).

\subsection{Detector response along Rindler trajectory}

We note that unlike in Fock quantization, two-point function (\ref{GPolyFinal}) 
\emph{does not} have any pole as $\eta\to1$. Therefore, the corresponding 
residue $\mathrm{Res}[f(\eta)]_{|\eta=1+i\delta}$ vanishes in polymer 
quantization. This in turns, leads \emph{non-transient} part of the induced
transition rate (\ref{RindlerGeneralR0}) to vanish \emph{i.e.}
\begin{equation}\label{FockRindlerR0}
R_{\omega}^0 = 0 ~. 
\end{equation}
Thus, from the response of the Unruh-DeWitt detector which interacts with 
polymer quantized scalar field, one would conclude that Unruh effect is not 
present in polymer quantization.
We should mention here that the large-distance two-point function including 
polymer corrections was computed in \cite{Hossain:2015xqa} and it may be
checked that \emph{Jordan's lemma} continues to be applicable even 
including polymer corrections.

As earlier, we may approximate the transient part of the transition rate by 
considering a reasonably large but finite observation time $\tau$. In 
particular, in the domain of perturbation where $ (a\lstar) \ll 1 $ and 
$e^{a\tau} \gg 1$, the transient part is  
\begin{equation}\label{IntegralDeltaRPolyRindler}
\Delta R_{\omega}({\tau}) \approx \frac{e^{-a\tau}[1+\mathcal{O}(a\lstar)]}
{\pi \left( 1+ \omega^2/a^2 \right)} 
\left[a \cos(\omega \tau) -\omega \sin(\omega\tau)\right].
\end{equation}
Clearly, the transient part of the induced transition rate decays exponentially 
in proper time $\tau$, similar to the results of Fock quantization.
We should mention here that the polymer quantization is known to cause a 
violation of Lorentz invariance due to the presence of the length scale 
$\lstar$. As one of the consequences, it is shown in 
\cite{Husain:2015tna,Kajuri:2015oza} that polymer vacuum state is not strictly 
invariant under the \emph{boost}. Therefore, the specific nature of the 
transient terms as studied here are effectively tied to the observer in polymer 
quantization.

\section{Discussions}

In summary, we have studied the properties of response function of an 
Unruh-DeWitt detector which moves along a Rindler trajectory and interacts 
weakly to a polymer quantized massless scalar field. Through a detailed 
calculation we have shown that unlike in Fock quantization, there are only 
transient terms present in the induced transition rate of the detector. In Fock 
quantization, the induced transition rate of the detector contains also  a 
non-transient term which is proportional to Planck distribution. This property 
of the induced transition rate signifies the existence of Unruh effect in Fock 
quantization. In polymer quantization of scalar field, however, the 
non-transient term is absent. Therefore, an Unruh-DeWitt detector would not 
perceive a flux of thermal particles along an uniformly accelerating trajectory 
in polymer quantization. The result as shown here provides an alternative 
evidence for the earlier reported results by the authors where it is shown using 
method of Bogoliubov transformation  \cite{Hossain:2014fma} as well as using 
Kubo-Martin-Schwinger (KMS) condition \cite{Hossain:2015xqa} that Unruh effect 
is absent in polymer quantization.

We note the key ingredients that led to the main result of this paper. As 
evident from the equation (\ref{RindlerGeneralR0}), the induced transition rate 
of an Unruh-DeWitt detector along a Rindler trajectory is quite generically 
proportional to the Planck distribution formula (\emph{i.e.} $\propto 
(e^{2\pi\omega/a} - 1)^{-1}$). However, the proportionality constant contains 
the residue of the two-point function evaluated at its pole. Therefore, the 
\emph{short-distance} singular nature of the Fock space two-point function is a 
key input to ensure the non-vanishing residue which in turns plays the crucial 
role for the existence of Unruh effect. On the other hand, it is widely expected 
that the short distance behaviour of the two-point function would receive 
significant modifications from possible Planck-scale physics. In this paper, we 
have shown that this expectation is indeed borne out in polymer quantization 
where the short-distance two-point function receives \emph{non-perturbative} 
modifications. In particular, the short-distance singular two-point function is 
replaced by a regular function which has no divergence in short-distance.

Finally, we may recall that several experiments have been proposed to 
detect possible signatures of Unruh effect in laboratory
\cite{Schutzhold:2006gj,Schutzhold:2008zza,Aspachs:2010hh}. Therefore, 
the result as shown here using response function of an Unruh-DeWitt 
detector and also shown earlier using methods of Bogoliubov transformation  
\cite{Hossain:2014fma} and Kubo-Martin-Schwinger (KMS) condition 
\cite{Hossain:2015xqa} where criticism raised in \cite{Rovelli:2014gva} was 
also addressed, clearly indicates that the experimental detection of 
Unruh effect can be used as a probe of possible Planck-scale physics. In 
particular, such a detection can be used to either verify or rule out a 
candidate Planck scale theory with a new dimension-full parameter, which affects 
quantization of matter fields as considered here. We conclude by acknowledging 
few similar recent results in the context of polymer quantization where it is 
shown that some high energy modifications can indeed lead to the alteration of 
certain low energy phenomena 
\cite{Husain:2015tna,Kajuri:2015oza}.
In particular, it is shown in \cite{Husain:2015tna,Kajuri:2015oza} that 
polymer vacuum state is not strictly invariant under the \emph{boost}. 
Therefore, the specific nature of the transient terms as studied here are 
effectively tied to the observer in polymer quantization. However, the transient 
terms are shown to decay out exponentially. So the detector response after a
reasonably long time, in the scale of $a^{-1}$, becomes essentially time 
independent. Clearly, the existence of Unruh effect needs to be understood from 
the properties of the non-transient part of the detector response as studied 
here.

\begin{acknowledgments}
We would like to thank Ritesh Singh for discussions. We thank Subhajit 
Barman and Chiranjeeb Singha for their comments on the manuscript. GS would 
like to thank UGC for supporting this work through a doctoral fellowship. 
\end{acknowledgments}

%\bibliographystyle{plain}
%\bibliographystyle{ieeetr}
%\bibliographystyle{apsrev}
%\bibliographystyle{aipauth4-1}

%\bibliography{bibtexfile}

\end{document}